\documentclass[11pt]{article}
\usepackage[dvips]{graphicx}

\begin{document}
\title{\Large \bf Isovector pairing in odd-A proton-rich nuclei} 
\author{ \\ Jonathan Engel  \\ 
{\small Dept.\ of Physics and Astronomy, CB3255, 
University of North Carolina, Chapel Hill, NC 27599  } 
\and  \\ Karlheinz Langanke \\ 
{\small Inst.\ for Phys.\ and Astr. and Cen.\ for Theor. 
Astrophys.,
Univ. of Aarhus, DK-8000 Aarhus C, Denmark }
\and  \\  Petr  Vogel \\
{\small Physics Department, California Institute of Technology, Pasadena, 
California 91125  }} 

%\date{\today} 
\maketitle
\begin{abstract} 
{\normalsize
A simple model based on the group SO(5) suggests that both the like-particle
and neutron-proton components of isovector pairing correlations in odd-$A$
nuclei are Pauli blocked.  The same effect emerges from Monte Carlo
Shell-model calculations of proton-rich nuclei in the full $fp$ shell.  There
are small differences between the two models in their representation of the
effects of an odd nucleon on the competition between like-particle and
neutron-proton pairing, but they can be understood and reduced by using a
two-level version of the SO(5) model.  On the other hand, in odd-odd nuclei
with $N \neq Z$, SO(5) disagrees more severely with the shell model because it
incorrectly predicts ground-state isospins.  The shell model calculations for
any $fp$-shell nuclei can be extended to finite temperature, where they show
a decrease in blocking.}
\end{abstract} 
%\pacs{21.10-k,21.60.Fw,21.60.Ka}

%\narrowtext
\newpage

Proton-rich nuclei play an important role in explosive nucleosynthesis and are
increasingly accessible to experiment.  For this reason the subject of
neutron-proton pairing has experienced a revival.  Even-$A$ nuclei have
received most of the attention so far, but odd-$A$ nuclei, in which pairing is
affected by the odd nucleon, are equally important.  In a previous
paper\cite{ELV} we analyzed the competition between neutron-proton ($np$) and
like-particle ($pp$ and $nn$) pairing in even-even $fp$-shell nuclei and in
odd-odd $N=Z$ nuclei, arguing that a simple model based on the group SO(5)
captured the essentials of full shell-model calculations, despite deformation,
spin-orbit splitting, and other physics that the simple model omits.  Here we
turn to odd-$A$ nuclei, discussing the same competition when an extra nucleon
is present and pairing correlations are blocked.  We also touch on odd-odd
nuclei, of which only those with $N=Z$ were treated in Ref.\ \cite{ELV}.
Though our focus is on ground states, we briefly discuss the pairing
competition at finite temperature as well.

We begin with the predictions of the simple model, a full description of which 
is in Ref.\ \cite{ELV}.  Briefly, the isovector angular-momentum-zero $nn$, 
$pp$, and $np$ pair creation operators, the corresponding annihilation 
operators, the three isospin generators, and the number operator form the 
algebra SO(5).  The Hamiltonian consists of three equally weighted pairing 
terms and has the ground-state expectation value
\begin{equation}
E=-\frac{G}{2}\left[ n(\Omega - \frac{n-6}{4}) - \nu(\Omega - \frac{\nu-6}{4}) 
+t(t+1)-T(T+1) \right]~,
\label{e:energy}
\end{equation}
where $G$ is a the pairing-force strength constant,
$n$ is the number of nucleons, $\Omega$ is half the number of (degenerate) 
single-particle levels, $T$ is the isospin, and $\nu$ and $t$ are the the 
seniority and ``reduced isospin", which take the values  $\nu =0$, $t=0$ in 
even-A ground states and $\nu =1$, $t=1/2$ in odd-A ground states.  When 
$\nu=0$ Eq.\  (\ref{e:energy}) becomes
\begin{equation}
E=-\frac{G}{2}\left[n (\Omega -\frac{n-6}{4} ) -T(T+1) \right]~.
\label{e:nu0}
\end{equation}
When $\nu=1$ we have
\begin{equation}
E=-\frac{G}{2}\left[n (\Omega -\frac{n-6}{4}) -(\Omega +\frac{1}{2})-T(T+1) 
\right]~.
\label{e:nu1}
\end{equation}

These equations imply a blocking in odd-$A$ nuclei very similar to what would
be present without $np$ pairing; in fact in the ordinary like-particle
seniority model the pairing energies along any even-$Z$ isotope chain,
in which there are no odd-odd nuclei, differ only by a small constant from
Eqs.\ (\ref{e:nu0}) and ( \ref{e:nu1}).  This prediction is confirmed by
shell-model Monte Carlo calculations, which we describe shortly.  Since the
$np$ pairing replaces some of the like-particle pairing when an
isospin-symmetric interaction is used, it would seem that both pairing modes 
are blocked in odd-$A$ nuclei, though $pp$ pairing is apparently
almost unaffected by an odd neutron.

This can be seen explicitly by examining the competition among $nn$, $pp$, and
$np$ pairs, something we discussed in even-$A$ nuclei in Ref.\ \cite{ELV}.  
Defining pair ``number operators" as in that paper, (for a brief but 
exhaustive discussion of these operators see Ref.\ \cite{Dobes}), we use 
techniques described in Ref.\ \cite{H} to obtain for odd-$A$ ground states:
\begin{eqnarray}                
\langle {\cal N}_{pp} \rangle &=&(T_c+1){({\cal N}-T_c) (1-\frac{{\cal 
N}-T_c-3}{2\Omega})} \over {2T_c+3} \nonumber \\
\langle {\cal N}_{np} \rangle &=&{({\cal N}-T_c) (1-\frac{{\cal 
N}-T_c-2}{2\Omega})} \over {2T_c+3} \nonumber \\                
\langle{\cal N}_{nn} \rangle &=&(T_c+1){({\cal N}+T_c) (1-\frac{{\cal 
N}+T_c+1}{2\Omega}) + (T_c+\frac{{\cal N}}{\Omega})} \over 
{2T_c+3} 
\end{eqnarray}
if ${\cal N} \equiv (n-1)/2$ (the total number of pairs) and $T_c \equiv 
T-1/2$ (the core isospin) are both 
even or both odd, and
\begin{eqnarray}
\langle{\cal N}_{pp} \rangle &=&T_c \, {({\cal N}-T_c)  (1 - 
\frac{{\cal N}-T_c-1}{2\Omega})} \over {2T_c+1} \nonumber \\	 
\langle{\cal N}_{np} \rangle &=&{({\cal N}-T_c+1) (1-\frac{{\cal  
N}-T_c-1}{2\Omega})} \over {2T_c+1} \nonumber \\	
\langle{\cal N}_{nn} \rangle &=&T_c \, {({\cal N}+T_c)  (1 - \frac{{\cal 
N}+T_c-3}{2\Omega}) + (T_c-1)(1-\frac{{\cal N}+T_c}{\Omega})} \over 
{2T_c+1}
\end{eqnarray}
if ${\cal N}$ and $T_c \equiv T+1/2$ are both even or both odd (the relation
between $T_c$ and $T$ is different in the two sets of expressions).  Combining
these results with those for even-$A$ nuclei from Ref.\ \cite{ELV}, we plot on
the left-hand side of Figure 1 the numbers of each kind of pair (scaled by
0.5; we discuss this factor shortly) as neutron number increases in the Cr
isotopes.  As the figure shows, $\langle {\cal N}_{nn} \rangle$ oscillates 
sharply while the
other pair-numbers stagger less.  The reason is that when $N$ is even as well
as $Z$, $nn$ and $pp$ pairing (particularly the former for $N>Z$) are enhanced
at the expense of $np$ pairing.  Adding a neutron to make $N$ odd blocks the
$nn$ pairing, thereby reducing $\langle {\cal N}_{nn} \rangle$, but also 
blocks $np$ pairing to a degree so that $\langle {\cal N}_{np} \rangle$ cannot 
take advantage of the reduced coherence in the $nn$ condensate.  A little 
surprisingly, perhaps, $\langle {\cal N}_{pp} \rangle$ is not able to take any 
advantage of the slight drop in $np$ pairing;
it doesn't increase until $N$ becomes even again and $\langle {\cal N}_{np} 
\rangle$ is reduced more significantly by the increased strength of the $nn$ 
condenstate.

\begin{figure}[htb]
\begin{center}
\includegraphics[width=12cm,height=10cm]{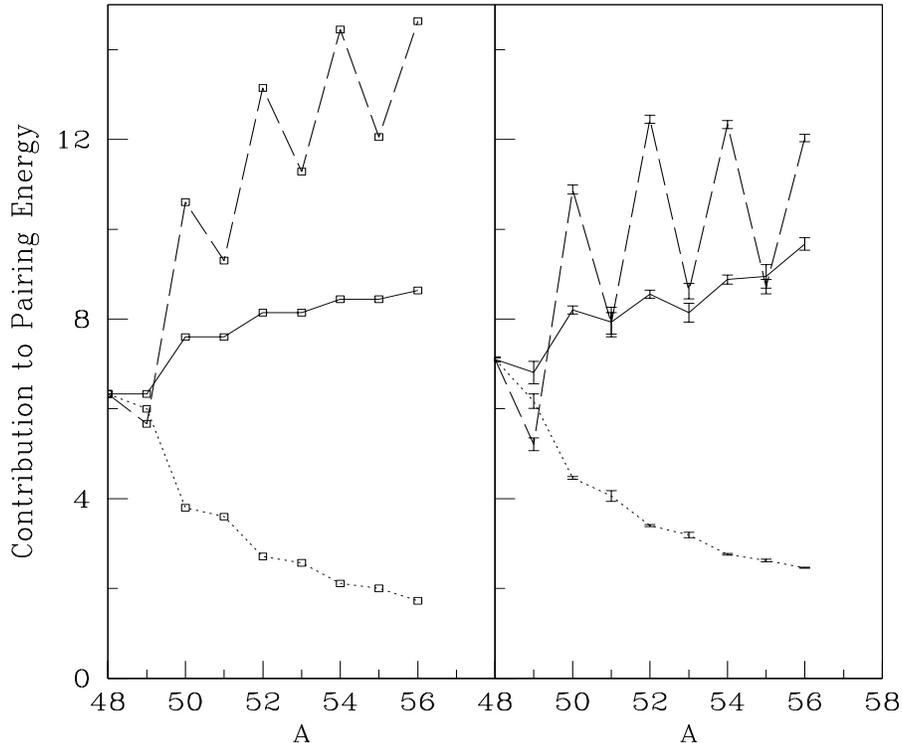}
\end{center}
\caption{The quantities $\Omega \langle {\cal N}_{pp} \rangle$ 
(solid line), $\Omega \langle {\cal N}_{nn} \rangle$ (dashed line), and 
$\Omega \langle {\cal N}_{np} \rangle$ (dotted line) for the Cr isotopes. On 
the left are the SO(5) results described in the text (with $\Omega = 10$, half 
the  number of single-particle levels in the $fp$ shell) scaled by a factor 
0.5.  On the right are the results of the Shell Model Monte Carlo 
calculation with $G_{\rm pair}= 20/A$ MeV and the quadrupole coupling constant 
$\chi = 134 A^{-11/3}$ MeV/fm$^4$. }
\end{figure}

This picture augments that described in Ref.\ \cite{ELV} and it is natural to
ask how much it has to do with reality.  Since the strengths of the individual
pairing modes and even the total isovector pairing strength are hard to
extract from the limited data available, we again turn to a large-scale shell
model Monte Carlo (SMMC) calculation in the full $fp$ shell to see if the
physics plays out in the same way.  Unfortunately in odd-$A$ and odd-odd
(excepting $N=Z$) nuclei the notorious sign-problem\cite{report} inherent in
SMMC studies with realistic interactions cannot be circumvented at low
temperatures ($T \leq 0.8$ MeV) by ``$g$-extrapolation"\cite{Alhassid}.  We in
large part avoid the problem, however, by using ``pairing plus
multipole-multipole" interaction of the kind used in Ref.\ \cite{Zheng}.  In
even-even nuclei this interaction has been shown to do a good job with
essential features of the spectrum, including isovector pairing correlations,
which have been checked against the predictions of the realistic KB3
interaction.  We fix the temperature at $T=0.4$ MeV, which should be
sufficient to cool a nucleus to near its ground state.  Although a mild sign
problem still affects the nuclei considered here (the sign for $^{49}$Cr is
$0.35\pm0.01$ at $T=0.4$ MeV) SMMC calculations can be performed without
$g$-extrapolation.  The only consequence of the residual sign problem is that
statistical uncertainties for odd-$A$ nuclei are slightly worse than for
even-even nuclei, in which there is no sign-problem at all.

The right-hand sides of Figure 1 contains the SMMC results for the same
quantities discussed in SO(5), and is clearly similar to the left-hand side.
The minor differences between the two panels, most apparent in $\langle {\cal
N}_{nn} \rangle$ for large $N-Z$, are most likely due to presence of several
nondegenerate levels in the $fp$ shell.  We conclude, as before, that the
presence of physics beyond SO(5) reduces the strength of each the three
pairing modes by about a factor of two, but does not drastically alter the
balance of power among them.

Where exactly does the factor of two come from?  The shell-model contains many
physical effects not included in SO(5), but we focus here on the role of
spin-orbit splitting, which we can mock up in a two-level version of the SO(5)
model.  (For the formalism, applicable to seniority-zero states, see Ref.\
\cite{Dussel,Civitarese}.)  Although the matrices one must diagonalize are no
longer tiny, they are still small.  To mimic the separation of the $f_{7/2}$
level from the rest of the $fp$ shell, we take our two levels to have
$\Omega_1=4$ and $\Omega_2=6$ and split them by $\epsilon=10 G$, a number
close to the real ratio of spin-orbit splitting to pairing-force strength.  In
Figure 2 we plot the numbers of the 3 kinds of pairs together with the SMMC
predictions, now just for even-A Cr isotopes (i.e.\ for every other point in
Figure 1).  The splitting weakens the coherence of the pairs so that the
numbers now agree quite well with the shell-model calculations without the
factor of 2 scaling, except for $\langle {\cal N}_{nn} \rangle$ at large
$N-Z$.  The disagreement there clearly reflects the residual splitting between
the other $fp$ levels; it doesn't appear until there are more neutrons than
can be accomodated in the $f_{7/2}$ level.  All this leads us to attribute the
differences in scale between simple SO(5) and full shell-model calculations to
spin-orbit splitting, or more generally to the nondegeneracy of the the single
particle states, which can also reflect deformation.

\begin{figure}[htb]
\begin{center}
\includegraphics[width=12cm,height=12cm]{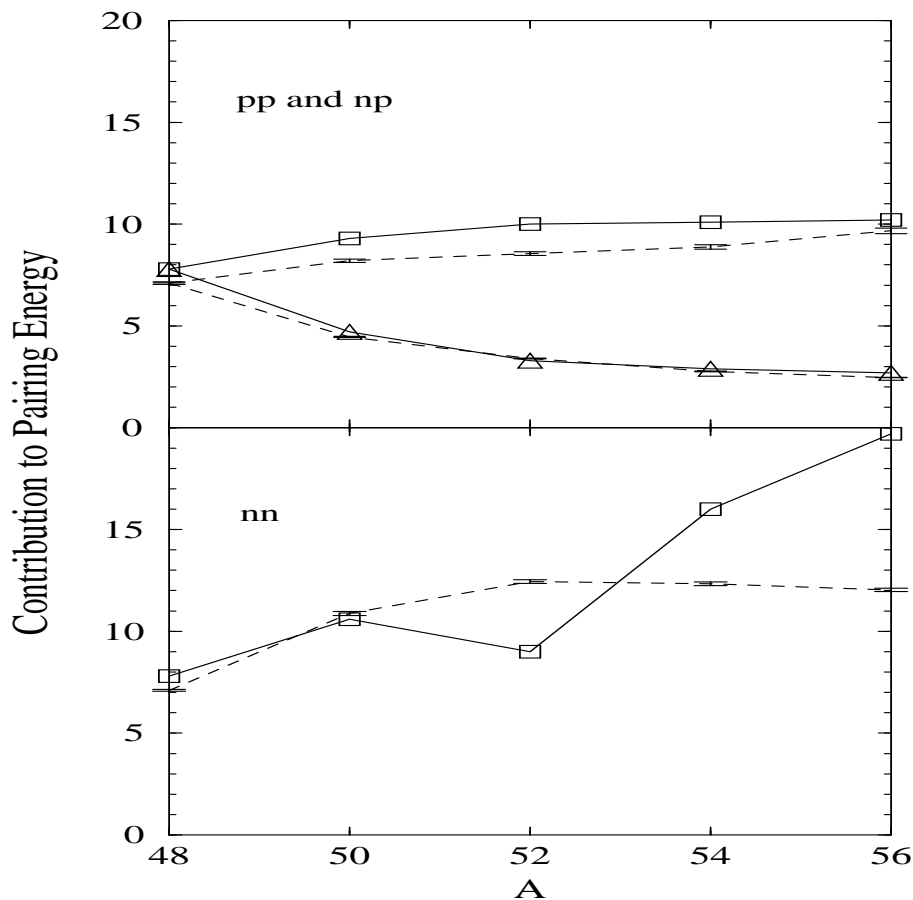}
\end{center}
\caption{Comparison of the two-level model and shell model.
In the upper panel are the quantities $\Omega \langle {\cal N}_{pp} \rangle$
and  $\Omega \langle {\cal N}_{np}\rangle$ and in the lower one
$\Omega \langle {\cal N}_{nn} \rangle$. The two level model is described in
the text; the results are connected by full lines. The SMMC results,
with the small error bars as indicated, are connected by the dashed lines.}
\end{figure}

We turn now to the Mn isotopes, which have $Z=25$ and are odd-odd as well as
odd-even.  Figure 3 shows SO(5) and SMMC results for the pair numbers.  The
figure is split into 4 panels to separate the odd-$A$ isotopes from the
even-$A$ (odd-odd) isotopes.  The agreement between (scaled) SO(5) and the
SMMC is good in the upper panels , but noticeably less so in lower panels.
The problem is that the model predicts the wrong isospin for the ground states
in odd-odd isotopes with $N-Z \neq 0$.  The ground states in all these nuclei
have $T=T_z$, while the model doesn't even contain such states; they appear
only when a pair is broken and, if you believe the model, should lie much 
higher in energy.  

\begin{figure}[htb]
\begin{center}
\includegraphics[width=12cm,height=10cm]{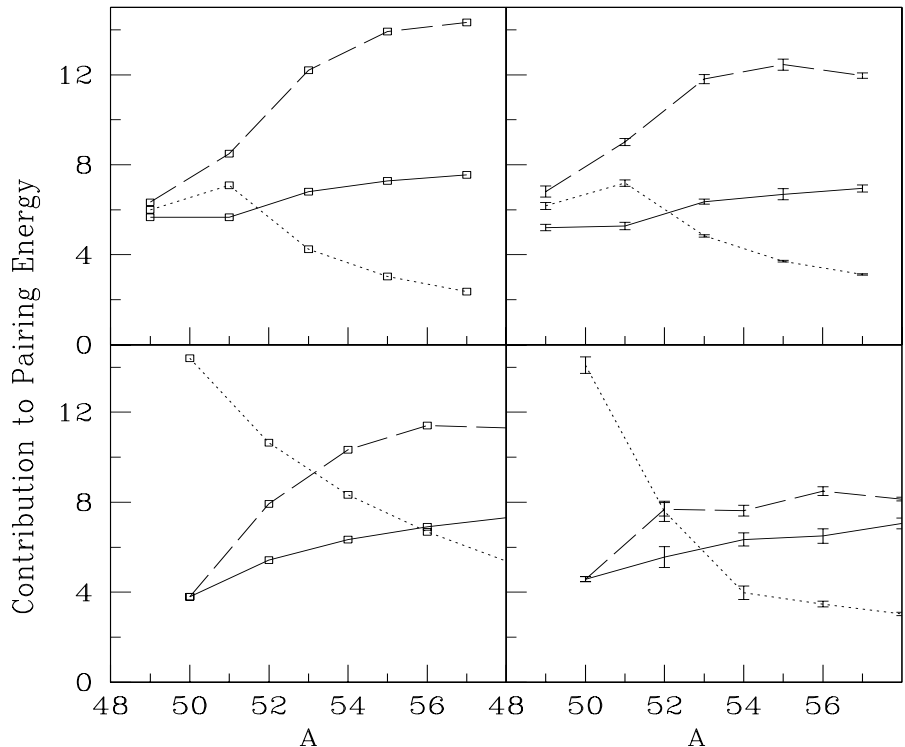}
\end{center}
\caption{The quantities $\Omega \langle {\cal N}_{pp} \rangle$ 
(solid line), $\Omega \langle {\cal N}_{nn} \rangle$ (dashed line), and 
$\Omega \langle {\cal N}_{np} \rangle$ (dotted line) in the odd-$A$ (top) and 
odd-odd (bottom) isotopes of Mn.  The left side contains the SO(5) results, 
the right the  SMMC results.}
\end{figure}

In the real world, however, the breaking of a pair is offset by the shell
closure at $N=28$ and by the quadrupole-quadrupole force, for the following
reason:  When $N \geq 28$ ($A \geq 53$ here) the lowest-lying states in the
pure single-particle model have $T=T_z$ because the filled neutron shell
prohibits the operator $\tau_+$ from giving anything but zero.  Higher $T$
states correspond to particle-hole excitations, which have too much
single-particle excitation energy to be pulled all the way down when the
pairing interaction is added.  A similar argument in the Nilsson scheme
implies that $T=T_z$ in the deformed nucleus $^{52}$Mn as well.  Our version
of SO(5) includes only fully paired states even in the two-level model, and so
overlooks the effects of single-particle splitting and deformation.  As a
result, its description of odd-odd nuclei with $N \neq Z$ is lacking.
\footnote{Interestingly, the SMMC with our chosen strength of the pairing plus
quadrupole interaction also fails to predict the correct ground-state isospin
in $^{52}$Mn.  The problem is clearly in the quadrupole-quadrupole force,
which is a little too weak.  Strengthening it by about 10\% pulls the
$T=T_z=1$ state below the $T=2$ state, as it is in reality.}

We turn finally to the temperature dependence of pairing correlations in
proton-rich nuclei (related work on nuclei with $N=Z$ nuclei appears in Refs.\
\cite{Zheng,Langanke96}).  As an example we have chosen $^{49}$Cr, an odd-$A$
nucleus with $N=Z+1$.  We again use the SMMC with the Hamiltonian
and model space described above to calculate pair-numbers, but now for
temperatures spanning the range $T=0.4 - 2$ MeV.  Figure 4 shows the results.
The blocking effect is clearly visible in the pair correlations:  at low
temperatures the neutron-neutron correlations are smaller than both the
proton-neutron and proton-proton correlations, despite the presence of five
neutrons to only four protons.  As the temperature increases both the pairing
strengths and the blocking effect decrease, so that for $T>1.2$ MeV the
neutron-neutron pairing correlations are the largest and the neutron-proton
the weakest, just as if there were no blocking.  It appears that although both
pairing correlations and Pauli blocking decrease with temperature, the latter
disappear first.

\begin{figure}[htb]
\begin{center}
\includegraphics[width=10cm,height=12cm,angle=-90]{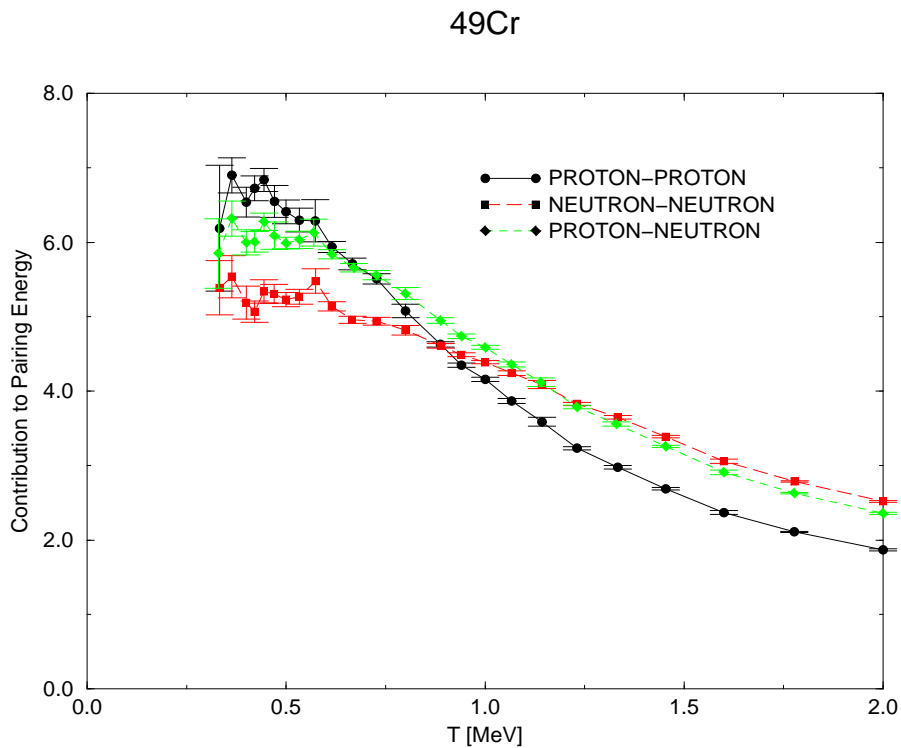}
\end{center}
\caption{The quantities $\Omega \langle {\cal N}_{pp} \rangle$ 
(circles), $\Omega \langle {\cal N}_{nn} \rangle$ (squares), and 
$\Omega \langle {\cal N}_{np} \rangle$ (diamonds) in the odd-$A$ 
nucleus $^{49}$Cr as a function of temperature calculated in the SMMC.}
\end{figure}

One can imagine using functional methods to study temperature-dependence in
SO(5) as well, provided some higher seniority states are included.  We defer
that exercise, however, and instead repeat in closing that although SO(5)
predicts the wrong isospin for odd-odd nuclei with $N \neq Z$, it makes the
competition among the three kinds of isovector pairs in all other isotopes
very easy to understand.  Most of the effects that escape the model are
recovered by a simple modification --- two sets of degenerate levels instead of
one --- even when Pauli blocking, which affects all three modes and the 
competition among them, is at play.

We were supported in part by the U.S.\ Department of Energy under grants 
DE-FG05-94ER40827 and DE-FG03-88ER-40397, by the U.S.\ National Science 
Foundation under grants PHY94-12818 and PHY94-20470, and by the Danish
Research Council.

\end{document}